\documentclass{article}
\usepackage[preprint]{spconf}
\usepackage{amsmath,graphicx}
\usepackage{amssymb}
\usepackage[hidelinks]{hyperref}
\usepackage{booktabs}
\usepackage{multirow}
\usepackage{cite}

\title{Using synthetic audio to improve the recognition of out-of-vocabulary words in end-to-end ASR systems}

\name{Xianrui Zheng$^{1*}$, Yulan Liu$^2$, Deniz Gunceler$^2$, Daniel Willett$^2$\thanks{$^*$Work done while interning at Amazon Alexa.}}
\address{$^1$University of Cambridge, $^2$Amazon Alexa}
\email{$^1$xz396@eng.cam.ac.uk, $^2$\{lyulan,denizg,dawillet\}@amazon.com}

\copyrightnotice{\copyright\ IEEE 2021}
                \toappear{To appear in {\it Proc.\ ICASSP2021,
                   June 06-11, 2021, Toronto,  Ontario, Canada
                   }}

\begin{document}

\maketitle

\begin{abstract}
Today, many state-of-the-art automatic speech recognition (ASR) systems apply all-neural models that map audio to word sequences
trained end-to-end along one global optimisation criterion in a fully data driven fashion.
These models allow high precision ASR for domains and words represented in the training material but
have difficulties recognising words that are rarely or not at all represented during training, i.e. trending words and
new named entities.
In this paper, we use a text-to-speech (TTS) engine to provide synthetic audio for out-of-vocabulary (OOV) words. We aim to boost the recognition accuracy of a recurrent neural network transducer (RNN-T) on OOV words by using the extra audio-text pairs, while maintaining the performance on the non-OOV words. Different regularisation techniques are explored and the best performance is achieved by fine-tuning the RNN-T on both original training data and extra synthetic data with elastic weight consolidation (EWC) applied on the encoder. 
This yields a 57\% relative word error rate (WER)
reduction on utterances containing OOV words without any degradation on the whole test set.
\end{abstract}

\noindent\textbf{Index Terms}: RNN-T, OOV words, synthetic audio by TTS

\section{Introduction}
\label{sec:introduction}
Traditional hybrid ASR systems consist of an 
acoustic model (AM), a language model (LM) and a pronunciation model (lexicon), 
all of which are trained independently. 
In contrast, all components are jointly trained in E2E ASR systems due to an integrated modelling structure. 
Some well-known E2E models include Listen, Attend and Spell (LAS)
\cite{chan2016listen} and RNN-T \cite{graves2012sequence}.

A challenge for these E2E models is that they require a large amount of labelled audio data for training to achieve good performance. 
For words that are not frequently seen in the training data (rare words) or 
not seen at all (OOV words), E2E models often find it difficult to 
recognise them \cite{Peyser2020}. 
Even though E2E models are typically trained to output sub-word tokens which 
can in theory construct some OOV words, in practice, words that are
OOV or rare in training suffer from recognition errors.
In case there is no correct hypothesis in the N-Best list or the lattice from the
beam search inference, it is also challenging  
for second pass methods such as LM rescoring \cite{Toshniwal2018} to improve recognition accuracy.
Hybrid ASR systems do not experience the same limitation, as the factorised components allow simpler updates
without the need for speech samples. The lexicon can be extended manually and
the LM can be updated with a small amount of targeted text data to support rare or OOV words.

In real-world voice assistant applications, it is common that an ASR system needs to predict 
rare or OOV words. 
For example, after the system is released, new trending words and named entities not 
included in the original training data might become important.
In a frequently encountered scenario for music applications, an ASR system needs to 
support new artists or newly published albums on an ongoing basis.
Furthermore, when extending functionalities of an assistant to new domains, it is highly likely
that existing training data does not cover the traffic in these new domains.
With hybrid ASR systems, such domain mismatch can be mitigated by updating the LM
with targetted text-only data. However, with E2E models, it is costly in terms of 
both time and financial resources to collect additional annotated audio data that 
contains rare and OOV words for each application domain.

Previous studies have improved the tail performance of
an E2E ASR system by combining shallow fusion with MWER fine-tuning \cite{Peyser2020},
or 
with a density ratio approach for LM fusion
\cite{McDermott2019}. 
These methods incorporate extra language models during decoding, thus increasing the amount of computation.
Few-shot learning strategies for E2E models are explored in \cite{Higy2018}, but 
in a small vocabulary command recognition task only. 

This paper focuses on
improving the performance of an existing word-piece based RNN-T model 
on new trending words which are completely missing in the training data, 
i.e. trending OOV words, without doing shallow fusion or second pass rescoring using an extra LM.
In particular, a TTS engine is used to generate audio from text data containing 
OOV words, and the synthetic data is used to improve the recognition accuracy for OOV words. 
Various regularisation techniques for fine-tuning are investigated and shown to be critical for both boosting the performance on OOV words and minimising the degradation on non-OOV words.

\section{Related Work}
\label{sec:related_work}
Domain adaptation is a relevant research thread that improves the performance of 
an ASR model on the test data following a different statistical distribution  
from the training data.
To tackle the domain mismatch with text-only data from the target domain, 
the output of E2E ASR models can be interpolated via shallow-fusion with an LM
trained on the target domain text data \cite{gulcehre2015using}. 
Another approach is to employ TTS to generate synthetic audio based 
on text data from the target domain.
The synthetic audio-text pairs can be used to adapt an E2E model 
\cite{sim2019personalization,li2020developing} to the target domain, 
or to train a spelling correction model \cite{guo2019spelling,li2020developing}.

Employing synthetic audio from TTS for ASR training recently gained
popularity as a result of advancements in TTS.
Recent research \cite{rosenberg2019speech, li2018training, laptev2020you}
has studied creating acoustically and lexically diverse
synthetic data, exploring the feasibility of 
replacing or augmenting real recordings 
with synthetic data during ASR model training, without compromising recognition performance. The results show 
that synthetic audio can improve the training convergence 
when the amount of available real data is as small as 10 hours but does not yet
replace real speech recordings to achieve the same recognition performance
given the same text sources \cite{rosenberg2019speech}.
In \cite{Leveraging2018}, instead of mapping texts to waveforms with an 
extra vocoder, the mel-spectrograms are synthesised directly and 
used to update the acoustic-to-word (A2W) attention-based sequence-to-sequence 
model \cite{chorowski2014end,chorowski2015attention}. \cite{Leveraging2018}
confirms that TTS synthetic data can be used to expand
the vocabulary of an A2W E2E model during domain adaptation. 
Another highlight from \cite{Leveraging2018,li2020developing} is that freezing all
encoder parameters is found beneficial 
when fine-tuning the model with synthetic data towards the target domain.

The mismatch in acoustic characteristics between real and synthetic audio can 
be problematic for ASR model training and fine-tuning.
Besides encoder freezing, another approach is to combine real and synthetic 
audio when fine-tuning the model,
which can also hinder catastrophic forgetting \cite{li2020developing}.
A third approach is to add a loss term to prevent the parameters of any adapted 
model from moving too far away from the baseline model. This approach is
particularly suitable for applications where the established domains covered by
the original training data and the target new domain are equally important.
The extra loss function can be as simple as the squared sum of the difference 
between the parameters prior to fine-tuning and during fine-tuning,
or it can be more advanced such as 
elastic weight consolidation (EWC) \cite{kirkpatrick2017overcoming}.

\section{Methodology}
\label{sec:methodology}
The E2E model used in this work is RNN-T
\cite{graves2012sequence,li2020developing,rnnt}.
Three methods are considered to fine-tune a baseline model and improve its performance
on new trending OOV words while maintaining the overall accuracy in the source domain. 

\subsection{Sampled data combination}
\label{subsec:method1_sample_all_data}

Fine-tuning on both synthetic and real data prevents the model from forgetting real audio. 
\cite{li2020developing} shows that a subset of the real data in the original source 
domain is needed if no regularisation method is applied during fine-tuning. 
In particular, \cite{li2020developing} kept about 30\% of source domain data used in training and combined them with the synthetic target domain data to form the final data for fine-tuning. 

Instead of directly combining 
a portion of the original real data with synthetic data, as in previous studies 
\cite{rosenberg2019speech,Rossenbach2019,li2020developing}, 
we propose to sample data on-the-fly from the source domain real data and 
the target domain synthetic data. 
The sampling distribution is a global and configurable hyperparameter
that propagates into each training batch. This allows a consistent sample mixing
and it also makes data combination independent from 
the absolute sizes of synthetic and real data.

\subsection{Encoder freezing}
\label{subsec:method2_encoder_freezing}

With encoder freezing (EF), the encoder parameters of a trained RNN-T are fixed, 
and only the parameters of the decoder and joint network are updated 
during fine-tuning.
In previous work \cite{li2020developing}, freezing the encoder 
provided much better results than not freezing the encoder
when fine-tuning RNN-T on synthetic data only.
This paper examines whether encoder freezing provides extra benefits 
on top of the sampled data combination method as explained in 
Section \ref{subsec:method1_sample_all_data}.

\subsection{Elastic weight consolidation}
\label{subsec:method3_l2_regularisation}

Encoder freezing only applies regularisation on the encoder in RNN-T,
but the unrealistic synthetic audio may indirectly negatively impact 
other components.
In addition, the word distribution of the text data for real recordings and synthetic 
audio is different.
When applying the method in Section \ref{subsec:method1_sample_all_data} during fine-tuning, the prior probability of words 
previously represented by real recordings 
is likely to decrease,
potentially causing a degradation in the overall 
WER of the source domain.
Since such changes in word probability are likely to impact the decoder and joint networks
more than the encoder, 
a regularisation for the decoder and joint networks may also be required during fine-tuning. 
We experiment with EWC \cite{kirkpatrick2017overcoming,y2020forget} for this 
purpose, with its loss function formulated as:

\begin{equation}
    \mathcal{L}_{\textnormal{EWC}} = \frac{\lambda}{2}\sum_i F_i(\theta_{\textnormal{new}, i} - \theta_{\textnormal{old},i})^2
    \label{eqn:ewc}
\end{equation}
where $\theta_{\textnormal{new}, i}$ is the current value for the $i$th parameter and $\theta_{\textnormal{old}, i}$
is the value of that same parameter before fine-tuning; therefore 
$\theta_{\textnormal{old}, i}$ is fixed throughout the fine-tuning process. 
$F_i$ is the diagonal entry of the fisher information matrix used to give the $i$th 
parameter a selective constraint.
$\mathcal{L}_\textnormal{EWC}$ is added to the regular RNN-T loss to force the 
parameters important to the source domain to stay close to 
the baseline model.

\section{Experimental setups}
\label{sec:experiments}
\subsection{Data}

We use 2.3K hours of anonymised far-field in-house data (Train19) as the training data of baseline RNN-T.
A dev set (Dev) and an eval set (Eval) are constructed with more recent data (1K hours each). 
Since Dev and Eval were from the live traffic of a later time period after Train19, some trending words in Dev and Eval may have rarely appeared in Train19. 

A list of OOV words is extracted by comparing Dev with Train19, i.e. 
all the words that have not appeared in Train19 but have appeared at least
three times in Dev. The minimal occurrence of three helps exclude typos from the
OOV word list. The utterances in Dev containing any OOV words are extracted
as a subset DevOOV, which contains 6.5K utterances and only accounts for 
0.7\% of the Dev set. 
Similarly, the utterances in Eval containing any OOV words
are extracted as a subset EvalOOV, containing 4.3K utterances. 
To reduce the decoding time, Dev and Eval are down-sampled randomly
to 200K utterances each into DevSub and EvalSub. The utterances
in EvalSub not covered by EvalOOV make another subset, i.e. EvalSub\textbackslash OOV, 
which is used to monitor the recognition accuracy of non-OOV words.

The US English standard voice of Amazon Polly is used to generate synthetic audio 
from the text of DevOOV with one voice profile. 
This TTS system is based on hybrid unit selection. Future work can use a better TTS to reduce acoustic mismatch between real and synthetic audio.

\subsection{RNN-T model}

A baseline RNN-T model is trained on Train19 until convergence. 
It has 5 LSTM layers in the encoder 
network (1024$\times$5), 2 LSTM layers in the decoder network (1024$\times$2), 
and a joint network with one feedforward layer (512$\times$1).
The output of the RNN-T is the
probability distribution of 4000 word-piece units from a unigram word-piece model \cite{kudo-2018-sp}.

\section{Results}
\label{sec:results}
We report the results in all tables with the normalised word error rate (NWER), which is the 
regular word error rate (WER) divided by a fixed number shared globally in this 
work, i.e. the WER of the baseline RNN-T on DevSub.
Each method in Section \ref{sec:methodology} was tested both on its 
own and in combination with other methods. 
The goal was to find the setup that gives the lowest WER on DevOOV without 
degrading the performance on DevSub, and the setup was further validated on 
EvalOOV and EvalSub.

\subsection{Fine-tune baseline model on combined data}

When fine-tuning the baseline RNN-T to improve the performance on OOV words, 
Table \ref{tab:TopOOVTTS} shows that it is important to control the sampling weights 
when combining Train19 and DevOOV. 
The model performs the worst on DevSub when fine-tuned on DevOOV with synthetic 
audio only, i.e. weights (0, 100).
As the percentage of samples from Train19 increases, the WERs on DevSub 
decreases, proving that combining synthetic data with the original training data 
can prevent the model from forgetting what it learnt before. 
Without degrading on DevSub, the best performance on DevOOV 
is observed with 70\% fine-tuning data from Train19 and 30\% from DevOOV with 
synthetic audio, 
achieving a 49\% relative WER reduction on DevOOV compared to the baseline.

\begin{table}[t]
\centering
\begin{tabular}{ c | c | r | r | r } 
\toprule
\textbf{Weights\%} & \multirow{2}{*}{\textbf{S/R}} & \multicolumn{3}{c}{\textbf{NWER}} \\
(R, S/R) & & DevOOV & DevSub & EvalOOV \\
\midrule
Baseline & - & 2.82 & 1.00 & 2.82 \\
\midrule
(0, 100) & S & 1.75 & 1.60 & - \\ 
(50, 50) & S & 1.37 & 1.02 & - \\ 
\textbf{(70, 30)} & \textbf{S} & \textbf{1.45} & \textbf{1.00} & \textbf{1.28} \\ 
(80, 20) & S & 1.50 & 1.00 & 1.34 \\
(90, 10) & S & 1.60 & 1.00 & 1.40 \\ 
\midrule
(70, 30) & R & 0.62 & 1.00 & 0.87 \\
(80, 20) & R & 0.10 & 0.99 & 0.32 \\
(90, 10) & R & 0.31 & 0.98 & 0.33 \\
\bottomrule
\end{tabular}
\caption{NWERs after fine-tuning the baseline on the combination of Train19 and DevOOV.
The weight on the left in the column \textbf{Weights\%} is the percentage of samples from Train19 and the weight on the right is the percentage of samples from DevOOV. \textbf{S/R} indicates whether real (R) or synthetic (S) audio is used to pair with DevOOV text data for fine-tuning.}
\label{tab:TopOOVTTS}
\end{table}

To find out the influence of the acoustic mismatch between real and synthetic audio, 
the last three rows of Table \ref{tab:TopOOVTTS} replace
synthetic audio with real recordings for DevOOV during fine-tuning. 
Fine-tuning on 
20\% of DevOOV with real audio achieves a 33\% relative WER reduction on EvalOOV compared to 
the same setup with synthetic audio. This  motivates the use of extra methods to
reduce the acoustic gap between real recordings and synthetic audio.

\subsection{Apply regularisation on the encoder}

Applying EF or EWC on the encoder may prevent the model 
from learning unwanted acoustic characteristics in synthetic audio.
Comparing Table \ref{tab:TopOOVTTS} with Table \ref{tab:TopTTSEF},
when fine-tuning on 100\% synthetic audio, 
applying EF gives 42\% and 19\% relative WER reductions on DevOOV and DevSub
respectively, and applying EWC on the encoder achieves a 
similar performance.
With weights (70, 30), i.e. the optimal setup from Table \ref{tab:TopOOVTTS}, 
EF improves the performance on DevOOV but degraded slightly on DevSub.
The degradation on DevSub can be recovered by increasing the percentage of
fine-tuning data sampled from Train19 to 90\%. 
Replacing EF with EWC
on the encoder further improves the WER on DevOOV by 2\% relative,
suggesting that freezing all encoder parameters during fine-tuning is suboptimal.
In this (90, 10) weights setup, compared to not applying any regularisation on the encoder in Table \ref{tab:TopOOVTTS}, EWC introduces 10\%, 1\% and 14\% relative WER reductions on DevOOV, 
DevSub and EvalOOV respectively.

\begin{table}[t]
\centering
\begin{tabular}{ c|c|r|r|r } 
\toprule
\textbf{Weights\%} & \textbf{Reg.} & \multicolumn{3}{c}{\textbf{NWER}} \\
(R, S) & E & DevOOV & DevSub & EvalOOV \\
\midrule
Baseline & - & 2.82 & 1.00 & 2.82 \\
\midrule
(0, 100) & EF & 1.02 & 1.30 & - \\ 
(0, 100) & EWC & 1.05 & 1.26 & - \\ 
\midrule
(70, 30) & EF & 1.26 & 1.02 & - \\ 
(80, 20) & EF & 1.33 & 1.01 & - \\ 
(90, 10) & EF & 1.47 & 0.99 & 1.22 \\ 
\midrule
\textbf{(90, 10)} & \textbf{EWC} & \textbf{1.44} & \textbf{0.99} & \textbf{1.21} \\ 
\bottomrule
\end{tabular}
\caption{Regularisation (\textbf{Reg.}) applied on encoder (E).}
\label{tab:TopTTSEF}
\end{table}
\vspace{-1mm}

\subsection{Apply regularisation on all components}

Table \ref{tab:TopTTSL2Encoder} adds EWC regularisation on the decoder and joint 
networks. 
EF is used instead of EWC on the encoder because EF does not require careful hyperparameter tuning and it performs similarly to EWC. 
Compared with Table \ref{tab:TopTTSEF}, 
when fine-tuning on DevOOV only with synthetic audio, regularising 
the decoder and joint networks with EWC helps mitigate the degradation on DevSub. 
However, when 90\% of fine-tuning data is sampled from Train19, 
adding EWC on the decoder and joint networks does not 
improve the performance on DevSub and introduces 7\% relative 
degradation on DevOOV.

\begin{table}[t]
\centering
\begin{tabular}{ c|c |c| r|r } 
\toprule
\textbf{Weights\%} & \multicolumn{2}{c|}{\textbf{Reg.}} & \multicolumn{2}{c}{\textbf{NWER}} \\
(R, S) & E & D, J & DevOOV & DevSub \\
\midrule
Baseline & - & - & 2.82  & 1.00 \\
\midrule
\textbf{(0, 100)} & \textbf{EF} & \textbf{EWC}  & \textbf{1.66} & \textbf{1.07} \\
(90, 10) & EF & EWC & 1.58  &  0.99 \\
\bottomrule
\end{tabular}
\caption{EWC applied on decoder network (D) and joint network (J) networks on top of EF.}
\label{tab:TopTTSL2Encoder}
\end{table}
\vspace{-1.1mm}

\subsection{Analysis on Eval subsets}

Two models are evaluated on subsets from Eval.
When the original training data is available, the model fine-tuned with the (90, 10) 
weights setup 
with EWC applied on the encoder (highlighted in Table \ref{tab:TopTTSEF}) is selected. 
Otherwise, we choose the model fine-tuned on 100\% synthetic audio with EF and 
EWC on the decoder and joint networks (highlighted in Table 
\ref{tab:TopTTSL2Encoder}). 
As shown in Table \ref{tab:TopOOVTTSeval}, having 90\% of fine-tuning data 
sampled from Train19 can almost eliminate the degradation on EvalSub\textbackslash OOV, 
leading to 57\% and 1\% 
relative WER reductions on EvalOOV and EvalSub respectively compared to the baseline.
In addition, for both models the improvement previously observed on DevOOV is 
successfully replicated on EvalOOV.

\begin{table}[t]
\centering
\begin{tabular}{ c|r|r|r } 
\toprule
\textbf{Weights\%} & \multicolumn{3}{c}{\textbf{NWER}} \\
(R, S) & EvalOOV & EvalSub\textbackslash OOV & EvalSub \\
\midrule
Baseline & 2.82 & 0.88 & 0.90 \\
\hline
(0, 100) & 1.37 & 0.96 & 0.96 \\ 
\textbf{(90, 10)} & \textbf{1.21} & \textbf{0.89} & \textbf{0.89} \\ 
\bottomrule
\end{tabular}
\caption{NWERs on EvalOOV, EvalSub\textbackslash OOV and Eval. }
\label{tab:TopOOVTTSeval}
\end{table}

\begin{figure}[t]
\centering
\includegraphics[width=\linewidth]{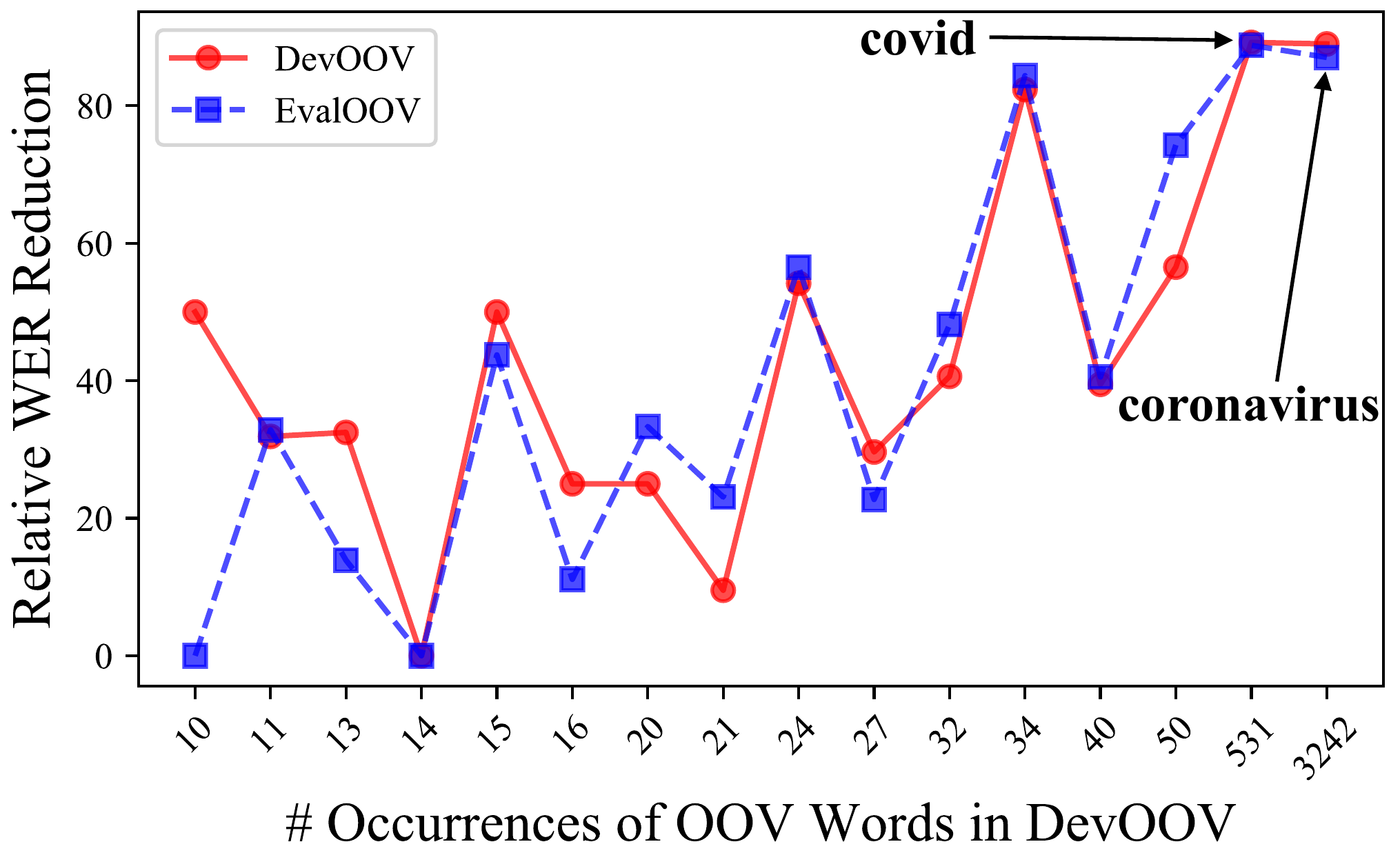}
\vspace{-7mm}
\caption{Relative WER reduction for OOV words.}
\label{fig:TOP_oov_line}
\end{figure}

Fig. \ref{fig:TOP_oov_line} shows the relative WER reduction averaged over all
OOV words that appear the same number of times in DevOOV, based on the
recognition results highlighted in Table \ref{tab:TopOOVTTSeval}. 
Overall, the recognition performance improves more if an OOV word is seen many 
times during fine-tuning, 
but the performance also improves for some OOV words that are seen only a few times.
The two most frequent OOV words are `coronavirus' and `covid'.
Even though the word-piece vocabulary can theoretically compose `coronavirus',
the baseline model only recognises one `coronavirus' in DevOOV. 
After using the best model in Table \ref{tab:TopOOVTTSeval}, the WER for 
`coronavirus' drops by 87\% relative on EvalOOV. 

\section{Conclusions}
\label{sec:conclusions}
This paper shows that using synthetic audio is an effective way to incrementally update an existing RNN-T model to learn OOV words. 
The best result gives a 57\% relative WER reduction on EvalOOV without degradation on EvalSub, indicating that the WERs of OOV words can be significantly reduced while preserving the WERs of non-OOV words. 
This is achieved by applying regularisation on encoder parameters and mixing the original training data with synthetic data during fine-tuning. 
Our study also shows that when fine-tuning on synthetic data only, applying regularisation on all RNN-T 
components can better mitigate the degradation in the overall WER than just applying 
regularisation on the encoder.

\bibliographystyle{IEEEtran}

\bibliography{bibliography}

\end{document}